\documentstyle[12pt,epsfig]{article}
\def\lsim{\raise0.3ex\hbox{$<$\kern-0.75em\raise-1.1ex\hbox{$\sim$}}}
\def\gsim{\raise0.3ex\hbox{$>$\kern-0.75em\raise-1.1ex\hbox{$\sim$}}}

\def\beq{\begin{equation}}
\def\eeq{\end{equation}}
\def\bea{\begin{eqnarray}}
\def\eea{\end{eqnarray}}
\def\bq{\begin{quote}}
\def\eq{\end{quote}}

\def\gappeq{\mathrel{\rlap {\raise.5ex\hbox{$>$}}
{\lower.5ex\hbox{$\sim$}}}}

\def\lappeq{\mathrel{\rlap{\raise.5ex\hbox{$<$}}
{\lower.5ex\hbox{$\sim$}}}}

\def\Toprel#1\over#2{\mathrel{\mathop{#2}\limits^{#1}}}

\begin{document}
\newcommand\ie {{\it i.e.}}
\newcommand\eg {{\it e.g.}}
\newcommand\etc{{\it etc.}}
\newcommand\cf {{\it cf.}}
\newcommand\etal {{\it et al.}}
\newcommand{\be}{\begin{eqnarray}}
\newcommand{\ee}{\end{eqnarray}}
\newcommand{\jp}{$ J/ \psi $}
\newcommand{\pp}{$ \psi^{ \prime} $}
\newcommand{\ppp}{$ \psi^{ \prime \prime } $}
\newcommand{\dd}[2]{$ #1 \overline #2 $}
\newcommand\noi {\noindent}
\pagestyle{empty}
\begin{center}
{\bf NUCLEAR MEDIUM EFFECTS IN $J/\Psi$ PRODUCTION AT HERA-B }
\\

\vspace*{1cm}
A. L. Ayala Filho  $^{1}$, C. Brenner Mariotto  $^{1}$ and    V.P. Gon\c{c}alves $^{1}$\\
\vspace{0.3cm}
{$^{1}$ Instituto de F\'{\i}sica e Matem\'atica,  Universidade
Federal de Pelotas\\
Caixa Postal 354, CEP 96010-090, Pelotas, RS, Brazil\\
$^{2}$ \rm
Departamento de F\'{\i}sica, Funda\c{c}\~ao Universidade Federal do Rio Grande \\
Caixa Postal 474, CEP 96201-900, Rio Grande, RS, Brazil}\\
\vspace*{1cm}
{\bf ABSTRACT}
\end{center}

\vspace*{1.5cm} \noindent

\vspace*{1.3cm} \noindent \rule[.1in]{17cm}{.002in}

\vspace{-3.5cm} \setcounter{page}{1} \pagestyle{plain}  

In this paper we estimate the influence of the shadowing effect and initial state parton energy loss in the quarkonium production at HERA-B. We analyze the $x_F$ behavior of the effective exponent $\alpha (x_F)$ and present a comparison with the preliminary HERA-B data for $J/\Psi$ production. Moreover, we estimate the magnitude of these effects in the $J/\Psi$ production at RHIC. 

\vspace{0.5cm}

The relativistic collider facilities RHIC and LHC  provide the opportunity to systematically study the
physics of hot and ultradense matter in hadron-nucleus ($pA$) and
nucleus-nucleus ($AB$) collisions at high energies (For a review see, e.g. \cite{hard_probes_cern,wang_jacobs}).
The systematic study of $pA$ collisions at the same energies is
essential to gain insight into the structure of the dense medium
effects. Such effects, as the energy loss and shadowing, are
absent or small in $pp$ collisions, but become increasingly
prominent in $pA$ collisions, and are of major importance in $AA$
reactions. By comparing $pA$ and $AA$ reactions involving very
heavy nuclei, one may be able to distinguish basic hadronic
effects that dominate the dynamics in $pA$ collisions, from a
quark-gluon formation predicted to occur in heavy ion $AA$
collisions. To gain insight into the underlying hadronic
processes, one has to study collisions that are expected to not
lead to a QGP formation. Once the physics of ``QCD at high
densities'' is better understood, the mechanisms of quark-gluon
plasma formation and related collective phenomena in heavy ion
collisions could be disentangled from the basic hadronic effects.

In this paper we study the influence of the nuclear medium effects in the quarkonium production,  particularly of the $J/\psi $,  which is one  of the
proposed signatures of the QCD phase transition \cite{satz}. In particular, we will consider the shadowing effects in the parton distributions and the initial state parton energy loss, which have the strongest influence on the  $x_F$ ($\equiv  x_1 - x_2$) behavior of the cross section (For a similar analyzes see, e.g. \cite{vogte866}).  Currently, the $A$ dependence of $J/\Psi$ production at $x_F > 0$ is known to rather high precision at several different energies (See e.g. \cite{ramona_rep}).
 On the other hand, the behavior of the cross section and the magnitude of the nuclear medium effects at negative $x_F$ region is still an open question. 
This situation should be improved by the experimental analyzes of quarkonium production in the fixed-target $pA$ experiment HERA-B at DESY, which measures the quarkonium $A$ dependence over -0.5 $< x_F <$ 0.3. First results for the $J/\Psi$ and $\Upsilon$  total cross sections have been recently published \cite{herab_total}. It  motivates a detailed study of the $A$ dependence for quarkonium production considering the current models for the nuclear medium effects. Here we  focus our analyzes in the HERA-B kinematical range considering two estimates for the magnitude of the initial state parton energy loss and two distinct parameterization for the shadowing effects. Moreover, we compare these predictions with the preliminary data recently obtained for the quarkonium production at negative values of $x_F$ \cite{dadosHERAB}. As we will show,  these data could be used to discriminate between the different models for the nuclear medium effects. As an extra possibility of discriminate the different effects, we also present the corresponding predictions for RHIC energies.

Lets start presenting a brief review about the nuclear medium effects.
One of the nuclear medium effects is the nuclear shadowing, which
is the modification of  the target parton distributions so that
$xq^A(x,Q^2) \, < \,  Axq^N(x,Q^2)$, as expected from a
superposition of $pp$ interactions (For a review see, e.g. \cite{reviews}). 
In the last  years several experiments have been dedicated to high precision
measurements of deep inelastic lepton scattering (DIS) off nuclei.
Experiments at CERN and Fermilab focus especially on the region of small
values of the Bjorken variable $x = Q^2/2M\nu$, where $Q^2=-q^2$ is the
squared four-momentum transfer, $\nu$ the energy transfer and $M$ the
nucleon mass. The data \cite{arneodo,e665}, taken over a wide kinematic
range, have shown that the proton and neutron structure functions are
modified by a nuclear environment. The modifications depend on the parton
momentum fraction: for momentum fractions $x < 0.1$ (shadowing region) 
and $0.3 < x < 0.7$ (EMC region), a
depletion is observed in the nuclear structure functions. These two regions are 
bridged by an enhancement known as antishadowing for $0.1 < x < 0.3$. We refer to the
entire phenomena as {\it the nuclear shadowing effect}.

The theoretical understanding of $F_2^A$ in the full kinematic region has
progressed in recent years, with several models which describe the
experimental data with quite success \cite{reviews}. Here we will
restrict ourselves to the descriptions which use the DGLAP evolution
equations \cite{dglap} to describe the behavior of the nuclear parton
distributions. In particular, Eskola, Kolhinen and Salgado (EKS)\cite{eks} have shown that the experimental results \cite{arneodo} presenting nuclear shadowing effects can be described using the
DGLAP evolution equations 
with adjusted initial parton distributions. The basic idea of this framework
is the same as in the global analyzes of parton distributions in the free
proton: they determine the nuclear parton densities at a wide range of $x$
and $Q^2\geq Q_0^2=2.25$ GeV$^2$ through their perturbative DGLAP evolution
by using the available experimental data from $lA$ DIS and Drell-Yan (DY)
measurements in $pA$ collisions as constraint. In this approach, the nuclear
effects are taken into account in the initial parton distribution $xf^A(x,Q_0^2)$ of the DGLAP evolution. EKS have expressed the
results in terms of the nuclear ratios $R_f^A(x,Q^2)$ for each parton flavor 
$f$ in a nucleus with $A$ nucleons ($A>2$), at $10^{-6}\leq x\leq 1$ and $
2.25\,GeV^2\leq Q^2\leq 10^4\,GeV^2$. Other groups have considered different set of data \cite{hkm} and/or next-to-leading order corrections to the DGLAP equation and proposed a distinct approach \cite{sassot} to describe the nuclear effects. In particular, De Florian and Sassot (DS) \cite{sassot} proposed a framework where each nuclear 
parton distribution is described by a convolution of the corresponding free nucleon parton distribution with a simple flavor dependent weight function that takes into account the nuclear effects. As pointed out by the authors,
the Mellin transform techniques allow a straightforward parameterization of the nuclear parton dynamical $Q^2$ evolution with a few parameters and an interpretation of the nuclear effects as a redistribution of longitudinal momentum among the partons in the nucleus.

\begin{figure}[t]
\centerline{
{\psfig{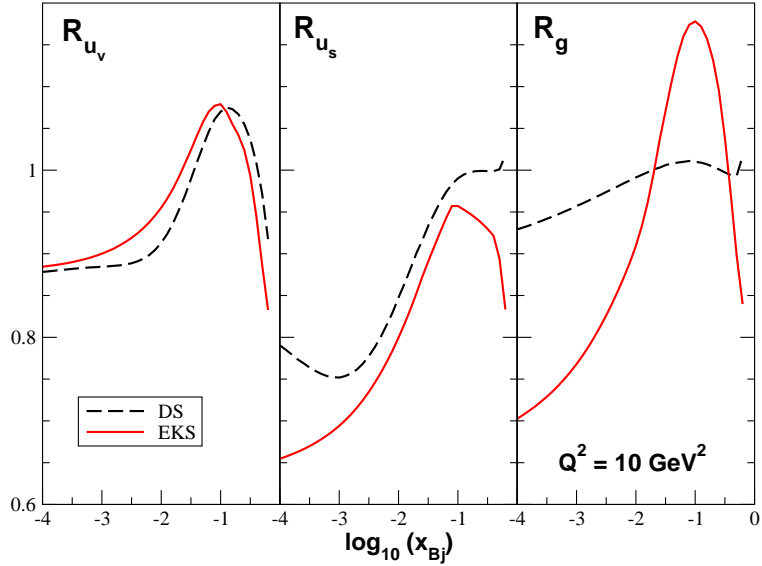}} }
\caption{Comparison between the EKS and DS parameterizations for the nuclear ratios $R_{u_V}^{184}(x,Q^2)$, $R_{u_S}^{184}(x,Q^2)$ and $R_g^{184}(x,Q^2)$.  }
\label{fig1}
\end{figure}

In Fig. \ref{fig1}  we present a comparison between the different parameterizations for the nuclear ratios $R_{u_V}^A(x,Q^2)$, $R_{u_S}^A(x,Q^2)$ and $R_g^A(x,Q^2)$ at $A = 184$. We can see that these parameterizations predict very distinct behavior for the nuclear parton distributions. In particular, the magnitude of the antishadowing in the nuclear gluon distribution is still an open question. This scenario should be improved by the experimental analyzes of  the quarkonium production at  HERA-B, since it probes the parton distributions in this $x$ range.

Another important effect in nuclear collisions is the initial state energy loss. In the last years the understanding of partonic energy loss has been extensively developed (For a review see e.g. \cite{hard_probes_cern,wang_jacobs}), because of the expectation that the order of magnitude of the effect in hot matter is much larger than in cold nuclear matter, which implies that the  resulting  jet quenching  can be considered a probe of the QGP formation. 
Our analyzes are focused in the parton energy loss in cold nuclear matter. It has been studied by Gavin and Milana \cite{GM} and
subsequently developed by Brodsky and Hoyer \cite{BH} and Baier et. al. \cite{BDMPS}, considering  a multiple scattering approach that essentially depletes the projectile parton momentum fraction, $x_1$, as the parton moves through the
nucleus.  The basic idea is that both the quarks and gluons can scatter elastically and loose energy
before the hard scattering. Consequently, the original projectile parton
momentum fraction $x_1$  when the parton first entered the target is modified to 
$x_1^{\prime} = x_1 - \Delta x_1$, where $\Delta x_1$ represents the loss in $x_1$ due to multiple
scatterings, being $x_1'$ the projectile parton momentum fraction involved in the hard scattering.  One has that the shifted value $x_1'$ enters the partonic cross
sections but the parton distributions must be evaluated at the initial $x_1$. Considering the relation 
between the averaged radiative energy loss $-dE/dz$ and  the characteristic squared transverse
momentum of the parton  $\langle p_{\perp W}^2 \rangle$, derived in  Ref. \cite{BDMPS} and  given by 
\be -\frac{dE}{dz} = \frac{3
\alpha_s}{4} \langle p_{\perp W}^2 \rangle \,\, \label{bdmpsdedz} \ee
one obtains that $\Delta x_1$ is
then \be \Delta x_1 = \frac{3 \alpha_s}{2} \frac{m_p}{x_1 s} L_A
\langle p_{\perp W}^2 \rangle \label{bdmpsdx} \ee
where $L_A$ is the nuclear medium length. As the average transverse momentum $\langle p_{\perp W}^2 \rangle$ is
proportional to $A^{1/3}$ \cite{BDMPS} (and $L_A\propto A^{1/3}$ as well), one has that 
$\Delta x_1 \propto A^{2/3}$ rather than $A^{1/3}$. In what follows we consider two estimates for $\langle p_{\perp W}^2 \rangle$ \cite{BDMPS}.  The larger value, which comes
from  single nuclear rescattering of photoproduced dijets estimated in Ref. \cite{LQS}, is given by  
\be  \langle p_{\perp W}^2 \rangle \simeq 0.658 \, \alpha_s \, A^{1/3} 
\, {\rm GeV}^2 \, \, . \label{bhmax} \ee  
Considering that the initial states could not be explicitly identified, one
assumes that $\langle p_{\perp W}^2 \rangle$ is identical 
for quarks and gluons.  In this case one has that 
when $\alpha_s \sim 0.3$ and $A=184$,  $-dE/dz
\simeq 1.28$ GeV/fm. We refer to this as ``LQS''
in the remainder of the discussion. 
 The second estimate takes into account the difference between quarks and gluon interactions and   has been derived in Ref. \cite{BDMPS} considering the relation between the    characteristic squared transverse
momentum of the parton and the  nucleon gluon distribution given by 
 \be \langle p_{\perp W}^2 \rangle_q & = & 
\frac{2 \pi^2 \alpha_s}{3} \rho_A xG(x,Q^2) L_A \simeq 0.07 \, \alpha_s \,
A^{1/3} \, {\rm GeV}^2 \label{bhminq}  \nonumber \\ 
\langle p_{\perp W}^2 \rangle_g & = & 
\frac{9}{4} \langle p_{\perp W}^2 \rangle_q \simeq 0.15 \, \alpha_s \,
A^{1/3} \, {\rm GeV}^2 \label{bhming} \ee
where $xG(x) \sim 1-2$ for the $x_1$ range of HERA-B. This lower estimate is
referred to subsequently as ``BDMPS''.  In this case one has that when $\alpha_s \sim 0.3$ and $A=184$, $-dE_q/dz \simeq 0.12$ 
GeV/fm and $-dE_g/dz \simeq 0.28$ GeV/fm.  
It is important to emphasize that a similar energy loss effect is expected for Drell-Yan production \cite{vogte866,arleo}. In Ref. \cite{arleo} the quark mean energy loss per unit length has been constrained to be $-dE_q/dz \simeq 0.2 \, \pm 0.15$ GeV/fm considering a leading order analyzes of E866/NuSea and NA3 Drell-Yan data, which reasonably agrees with the BDMPS estimate. 
A comment is in order here. As pointed out in Ref. \cite{vogte866}, the application of the BDMPS model at $x_F <0$ may becomes problematic since $\Delta x_1$ grows larger than $x_1$, suggesting that the calculation may not be applicable for $\Delta x_1 > x_1$. As it still is an open question, in a first approximation we will apply the model in the whole $x_F$ range.

In order to analyze the quarkonium production we will consider the  color
evaporation model (CEM) \cite{Mariotto:2001sv}. In this model, 
the color charges of the $c\overline c$ produced are randomized by the exchange of soft gluons, such that no information remains of the color configuration given by the preceding hard interactions. SU(3) algebra gives the probability  $1/9$ for the $c\overline c$ to be in a color singlet state and $8/9$ to be in a color octet state. It is then assumed that all color singlet pairs with invariant mass below the threshold for open charm will form a charmonium state. 
The cross section for for the charmonium state $i$ is then \be \frac{d \tilde{\sigma}_i}{d x_F} = \frac{\rho_i}{9}
\int_{2m_c}^{2m_D} \, dm_{c\overline c} \,  \frac{d \sigma^{c \overline c}}{dx_F
dm_{c\overline c}}
 \, \, , \label{cevap} \ee  
where $m_{c\overline c}$ is the invariant mass of the $c\overline c$ pair,
$m_{c}$ is the charm quark mass and $2m_D = 3.74$ GeV is the $D\overline D$ threshold. $\frac{d \sigma^{c \overline c}}{dx_F
dm_{c\overline c}}$ is the usual convolution of the perturbative QCD cross
section with the parton density functions for the proton/nucleus.
$\rho_i$ are nonperturbative universal factors which give the relative rates of producing the different charmonium
states. 
Once $\rho_i$ has been determined for each state, {\it e.g.}\ $\psi$, $\psi'$ or $\chi_{cJ}$, the model successfully predicts the energy and momentum dependencies.  
We note that $\rho_{\psi}$ includes both direct
$\psi$ production and indirect production through radiative decays of the
$\chi_{cJ}$ states and hadronic $\psi'$ decays. 
The pQCD cross section is taken in LO for simplicity, since we are just interested in ratios of cross sections. Also, the dependency on the overall factors ${\rho_i}/{9}$ cancels out in the ratios. It is important to emphasize that although we will use the CEM to describe the quarkonium production, similar results are expected if the nonrelativistic QCD model (NRQCD) \cite{NRQCD} is considered.

\begin{figure}
\centerline{
\begin{tabular}{ccc}
{\psfig{figure=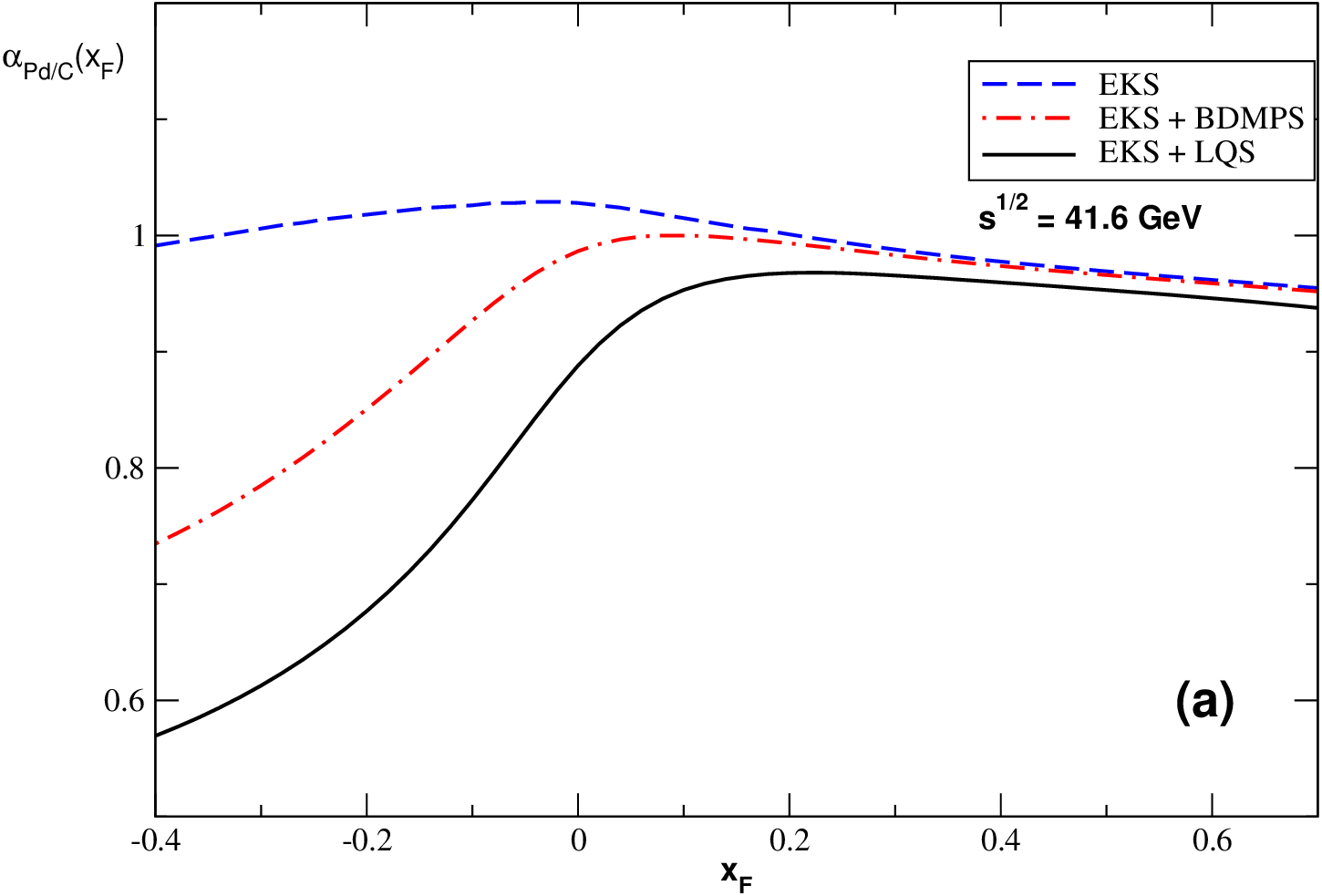,width=8.0cm}} & \,\,\,\,\,\, &
{\psfig{figure=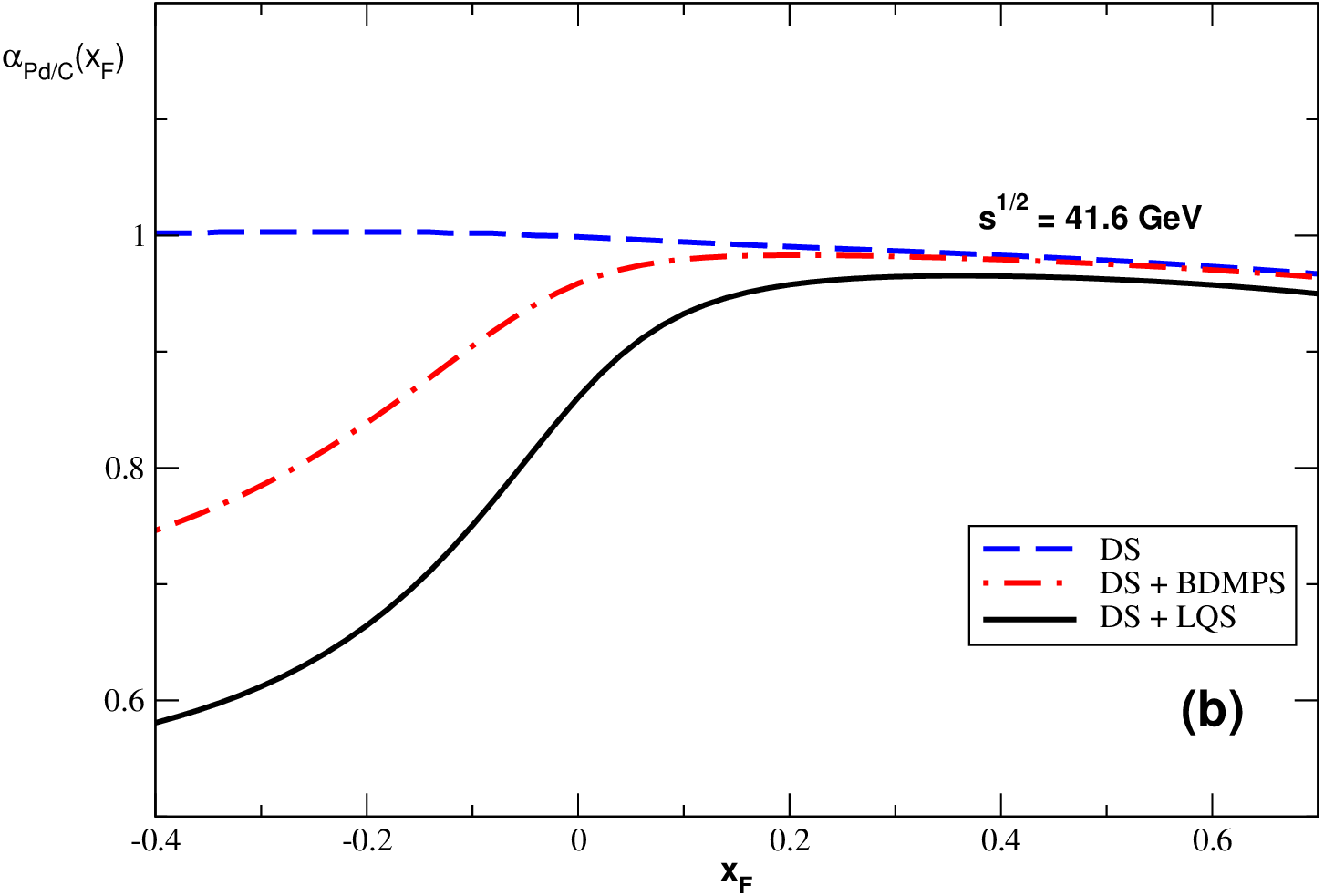,width=8.0cm}}\\
 & & \\
& & \\
\end{tabular}}
\caption{  Effective exponent as a function of $x_F$ for  charmonium production at HERA-B  considering  paladium and carbon
targets.}
\label{fig2}
\end{figure}


In order
to investigate the medium dependence of the quarkonium production
cross section, we will follow the usual procedure  used  to
describe the experimental data on nuclear effects in the hadronic
quarkonium production \cite{alfass}, where the atomic mass number
$A$ dependence is parameterized by $\sigma_{pA} = \sigma_{pN}
\times A^{\alpha}$. Here $\sigma_{pA}$ and $\sigma_{pN}$ are the
particle production cross sections in proton-nucleus and
proton-nucleon interactions, respectively. If the particle
production is not modified by the presence of nuclear matter, then
$\alpha = 1$. A number of experiments have measured a less than
linear $A$ dependence for various processes of production, which
indicates that the medium effects cannot be disregarded (See e.g. \cite{ramona_rep}). To estimate the modification of quarkonium production cross
section due to the medium effects, we calculate the
effective exponent $\alpha (x_F)$, which is given by
\begin{eqnarray}
\alpha(x_F) =  \left\{ \left. \ln \left(\left.{\frac{d
\sigma_{pA}}{d x_F}}\right/ {\frac{ d \sigma_{pN}}{d x_F}}
\right)\right/ {\ln A}\right\}\,\,. \label{alfaxf}
\end{eqnarray}
Moreover, to obtain the available observable measured in the experiments, we also replace the 
nucleon by a light nucleus target (Carbon),  
calculating the ratio 
\begin{eqnarray}
\alpha_{A_2/A_1}(x_F) =  \left\{ \left. \ln \left(\left.{\frac{d
\sigma_{pA_2}}{d x_F}}\right/ {\frac{ d \sigma_{pA_1}}{d x_F}}
\right)\right/ {\ln (A_2/A_1)}\right\}\,\,, \label{alfaxf2}
\end{eqnarray}
where $A_1 = C$ and $A_2 = Pd, W$.  
Our results for the effective exponent $\alpha_{A_2/A_1}(x_F)$ in $J/\Psi$
production at HERA-B energy of $\sqrt{s}=41.6$ GeV, with paladium and carbon
targets are shown in Fig. \ref{fig2} considering the EKS [Fig. \ref{fig2} (a)] and DS  [Fig. \ref{fig2} (b)] parameterizations for the nuclear shadowing effects. In the dashed curves only the nuclear shadowing 
effects are taken into account. Energy loss effects are
also included in the dot-dashed (BDMPS estimate) and full (LQS estimate) curves. The $x_F$ behavior of the  effective exponent when considering the EKS and DS parameterization is similar.
Whereas the shadowing alone produces a small enhancement for negative $x_F$
(antishadowing indeed) and a small suppression for higher positive values of
$x_F$, the inclusion of energy loss leads to a large suppression for the heavy nuclear
target, for negative values of $x_F$. This suppression can be understood in terms of the basic properties of the energy loss models. For a fixed value of negative $x_F$, the corresponding value of $x_1^{\prime}$ is smaller than the value of $x_1$ that enters in the parton distribution evaluation, since the projectil parton looses its momentum when traversing the nuclear target. This effect corresponds to a shift in the parton momentum to higher values of $x$ in the nuclear case, which reduces the amount of partons in the initial state of partonic subprocesses as compared to the free nucleon scattering, since the parton distributions grow as $x$ goes to zero due to the dynamical QCD evolution.

\begin{figure}
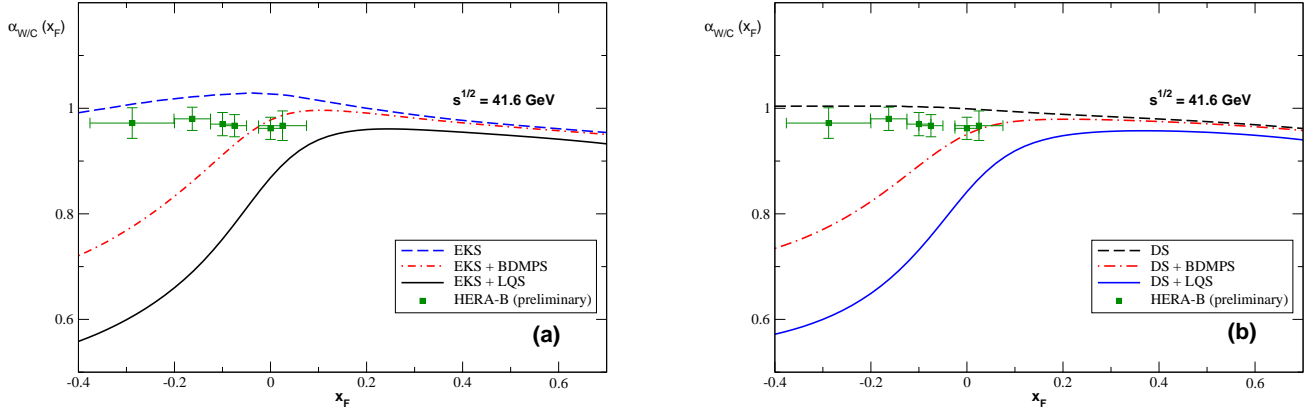

\centerline{
\begin{tabular}{ccc}
{\psfig{figure=herabWC2.eps,width=8.0cm}} & \,\,\,\,\,\, &
{\psfig{figure=herabWCfs2.eps,width=8.0cm}}\\
 & & \\
& & \\
\end{tabular}}
\caption{ Effective exponent as a function of $x_F$ for  charmonium production at HERA-B  considering  tungsten and carbon
targets.}
\label{fig4}
\end{figure}


In Fig. \ref{fig4} we present our  results for $J/\Psi$
production at same energy, in a tungsten and carbon targets, compared with
preliminary HERA-B data for $J/\Psi$ production \cite{dadosHERAB}. 
Concerning the EKS parameterization [Fig. \ref{fig4} (a)], neither the enhancement due to the shadowing effects alone, nor the strong 
suppression for $x_F<0$ due to energy loss are seen in the data, which indicate
a small and  $x_F$-independent suppression. These results may indicate that the magnitude of the antishadowing is overestimated in the EKS parameterization or that the energy loss is smaller that predicted by the LQS and  BDMPS models. When the shadowing effects are taken into account via DS parameterization, the prediction get closer to the HERA-B data. This is due to the fact that the DS parameterization does not predict anti-shadowing behaviour of the sea quark and gluon nuclear distributions, as we can see in Fig.\ref{fig1}. Again, when compared to the HERA-B data, the models BDMPS and LQS for the energy loss overestimate the suppression of the nuclear cross section for negative $x_F$.

\begin{figure}[t]
\centerline{
{\psfig{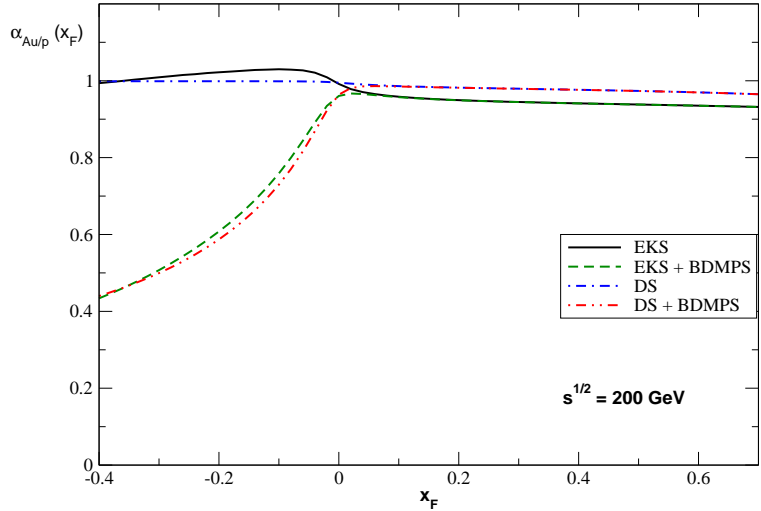}} }
\caption{Effective exponent as a function of $x_F$ for  charmonium production at  RHIC considering  gold  and proton targets. }
\label{fig5}
\end{figure}

Another possibility to constrain the magnitude of the medium effects is the study of the quarkonium production in proton(deuteron)-nucleus collisions at RHIC (See, e.g. Refs. \cite{vogtrhic,tuchin}). In this case, due to the larger center of mass energy, the  nuclear effects should be amplified. In Fig. \ref{fig5} we present our estimates for the effective exponent for $J/\Psi$ production at $\sqrt{s}=200$ GeV RHIC energy, with proton and gold targets. As we can see, the effective exponent is almost $x_F$ independent in the positive $x_F$ range, with the EKS prediction being smaller due to the larger shadowing present in this parameterization. On the other hand, the energy loss leads to a very strong suppression for negative $x_F$, larger than that predicted at HERA-B. It implies that the experimental analyzes of the quarkonium production  is ideal to constrain the initial state energy loss effects in cold nuclear matter.

One effect which we have disregarded in our analyzes was the nuclear absorption associated with the fact that 
the \dd{c}{c} pair may interact with nucleons and be dissociated or
absorbed before it can escape the target. This effect has been estimated in Ref. \cite{ramonaNPA700} considering different models for the quarkonium production and color singlet and color octet absorption. At HERA-B energy, the CEM and color singlet absorption in $J/\Psi$ production implies that $\alpha(x_F = -0.5) \approx 0.97$ and $\alpha (x_F > 0)  \approx 1$. On the other hand, if color octet absorption is considered one has $\alpha(x_F = -0.5) \approx 0.96$ and $\alpha (x_F > 0)  \approx 0.95$. Consequently,  if only the absorption effect is included in the calculations  a reasonable description of the preliminary data is possible.  However, the inclusion of this effect in combination with shadowing and energy loss effects 
will implicate a larger suppression at
negative $x_F$, which is disfavoured by the data. Thus, the estimate of the nuclear effects in quarkonium production is still an open question.

As a summary, in this paper we have studied the quarkonium production at HERA-B and RHIC. In particular, we have considered  two distinct parameterizations for the shadowing and  two estimates for the magnitude of initial energy loss effects and analyzed the $x_F$ behavior of the effective expoent. Our main emphasis was in the negative $x_F$ range which have been probed at HERA-B. We have verified that the inclusion of the energy loss strongly modify the behavior of $\alpha (x_F)$ in this kinematical range. The comparison of our predictions with the preliminary data indicates that the combination of shadowing and energy loss effects, as described by the EKS or DS parameterizations and LQS or BDMPS models, is not able to describe the data.  This may be
related to the applicability of the BDMPS approach in cold matter and/or the overestimation of the antishadowing effect in the EKS parameterization.
 Another possibility is that the correct model for the initial state parton energy loss is one similar to the Gavin-Milana model \cite{GM}, where $\Delta x_1 = \epsilon_i x_1 A^{\frac{1}{3}}$ ($\epsilon_q = 0.00412$ and $\epsilon_g = \frac{9}{4} \epsilon_q$), which implies $\alpha (x_F) \approx 1$ at negative $x_F$.   Another uncertainty present is the validity of the collinear factorization in nuclear collisions. Some authors advocate that the coherence phenomena cannot be disregarded \cite{kope}. Since the suppression predicted in RHIC is larger, it would be desirable to have 
measurements in that region to constrain the different models and disentangle
the different effects. To conclude, nuclear collisions remain a fascinating place to study different 
effects including the interplay of perturbative and non-perturbative QCD, and
nuclear effects.

\section*{Acknowledgments}
The authors are grateful to  R. Sassot for useful discussions.
This work was  partially financed by the Brazilian funding
agencies CNPq and FAPERGS.

\end{document}